\documentclass{article}

\usepackage{arxiv}

\usepackage[utf8]{inputenc} 
\usepackage[T1]{fontenc}    
\usepackage{hyperref}       
\usepackage{url}            
\usepackage{booktabs}       
\usepackage{amsfonts}       
\usepackage{amsmath}
\usepackage{nicefrac}       
\usepackage{microtype}      
\usepackage{graphicx}
\usepackage{natbib}
\usepackage{float}
\usepackage{doi}
\usepackage{xcolor}
\definecolor{EBblue}{RGB}{0,90,180}
\definecolor{TensionOrange}{RGB}{220,110,0}
\definecolor{TimoRed}{RGB}{190,30,45}
\definecolor{TensionGreen}{RGB}{0,200,70}
\definecolor{TimoGreen}{RGB}{0,150,70}

\begin{document}
\title{Competition between tension and bending waves in a pre-stretched soft beam : experimental validation}


\author{ \href{https://orcid.org/0000-0000-0000-0000}{\includegraphics[scale=0.06]{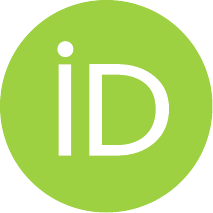}\hspace{1mm}Jiayuan Zhu} \\
	LabTAU, INSERM, Centre Léon Bérard, Université Claude Bernard Lyon 1, F-69003, LYON, France\\
	\And
	\href{https://orcid.org/0000-0000-0000-0000}{\includegraphics[scale=0.06]{orcid.pdf}\hspace{1mm}Jorge Torres} \\
    LabTAU, INSERM, Centre Léon Bérard, Université Claude Bernard Lyon 1, F-69003, LYON, France
    \And
	\href{https://orcid.org/0000-0000-0000-0000}{\includegraphics[scale=0.06]{orcid.pdf}\hspace{1mm}Sibylle Grégoire} \\
    LabTAU, INSERM, Centre Léon Bérard, Université Claude Bernard Lyon 1, F-69003, LYON, France
        \And
	\href{https://orcid.org/0000-0000-0000-0000}{\includegraphics[scale=0.06]{orcid.pdf}\hspace{1mm}François Legrand} \\
	LabTAU, INSERM, Centre Léon Bérard, Université Claude Bernard Lyon 1, F-69003, LYON, France \\
	Now at: Aix-Marseille Univ., CNRS, Centrale Med,Laboratoire de Mécanique et d’Acoustique, Marseille, France \\
        \And
	\href{https://orcid.org/0000-0000-0000-0000}{\includegraphics[scale=0.06]{orcid.pdf}\hspace{1mm}Frédéric Turquier} \\
    LBA, Aix Marseille Université, CNRS, Centrale Marseille, F-13013, Marseille, France\\
            \And
	\href{https://orcid.org/0000-0000-0000-0000}{\includegraphics[scale=0.06]{orcid.pdf}\hspace{1mm}Rohit Madke} \\
    LBA, Aix Marseille Université, CNRS, Centrale Marseille, F-13013, Marseille, France\\
            \And
	\href{https://orcid.org/0000-0000-0000-0000}{\includegraphics[scale=0.06]{orcid.pdf}\hspace{1mm}Vít Nováček} \\
    NTC, University of West Bohemia in Pilsen, Univerzitní 8, 301 00 Pilsen, Czech Republic. \\
                \And
	\href{https://orcid.org/0000-0000-0000-0000}{\includegraphics[scale=0.06]{orcid.pdf}\hspace{1mm}Stefan Catheline}\thanks{corresponding author} \\
    LabTAU, INSERM, Centre Léon Bérard, Université Claude Bernard Lyon 1, F-69003, LYON, France \\
	\texttt{stefan.catheline@inserm.fr} \\
                \And
	\href{https://orcid.org/0000-0000-0000-0000}{\includegraphics[scale=0.06]{orcid.pdf}\hspace{1mm}Bruno Giammarinaro} \\
    LabTAU, INSERM, Centre Léon Bérard, Université Claude Bernard Lyon 1, F-69003, LYON, France \\
}

\maketitle

\date{\today}

\begin{abstract}
Pre-stretched one-dimensional soft beams serve as an important and simplified model in mechanics, biomechanics, and soft material characterization. The propagation characteristics of transverse waves excited on such beams are commonly used to characterize elastic stiffness and tensile state. However, conventional pure beam or string models cannot fully explain the phase velocity dispersion of transverse waves in pre-stretched beams, particularly when the dispersion coexists with an increase in wave speed as tension increases. This frequency-dependent transition between tension-dominated and bending-dominated behavior has not yet been directly validated using measured dispersion curves in one-dimensional soft beams. Here, we use optical elastography, a high spatiotemporal resolution technique, to track transverse wave propagation in a soft incompressible cylinder under different tension levels and obtain the corresponding phase velocity dispersion curves. Through model-based inversion of the experimental dispersion curves, we demonstrate that a coupled Timoshenko-beam/string model successfully captures both the string-dominated and bending-dominated regimes observed experimentally. This result indicates that transverse wave propagation in pre-tensioned beams should be understood as a frequency-dependent competition between tension and bending effects. 
\end{abstract}

\section*{Introduction}
Many soft, beam-like structures, including tendons \citep{Martin2018GaugingForceTendons,Keuler2019AchillesTendonLoading}, ligaments \citep{Rougereau2022DeltoidLigamentSWE, Yancey2023SpinalLigamentSWE}, and spider silk \citep{ZHAO2021111208}, are subjected to axial loading or pre-stretched.  This is a boundary condition that plays an important role in guided-wave propagation.
When an axial pre-tension is applied, the transverse wave dynamics are governed by the competition between bending stiffness and tension, leading to tension-dependent wave speed and dispersion~\citep{graff2012wave}.  

These propagation regimes have already been studied using beam theory in different systems, including  soft rods \citep{salehabadiQuantifyingUniaxialPrestress2023} and  \textit{in vivo}  tendons  \citep{Martin2018GaugingForceTendons}.   \citet{salehabadiQuantifyingUniaxialPrestress2023} presented a propagation equation based on the tensioned Timoshenko model but did not extensively investigate the propagation because of the difficulty of retrieving the mechanical parameters through inversion. \citet{Martin2018GaugingForceTendons}, in the contrast, presented wave speed estimates using the same model. Their analysis considered only a high-frequency approximation, under which the dispersion of flexural waves is no longer retained. 

Flexural waves in the presence of pre-stress have also been investigated using elasticity theory coupled with incremental deformation theory \citep{ogdenNonlinearElasticDeformations1997}. Within this framework, \citet{ZHAO2021111208} demonstrated the possibility of retrieving the tension wave, also referred to as the string-like regime. The formulation, however, did not yield an equation combining tension and flexural-wave effects simultaneously.  This combination was introduced through a modified Rayleigh beam model. 
More recently, \citet{Delory2023HighlyStretchedPlate}  developed a semi-analytical finite element approach to retrieve different propagation modes in highly stretched soft plates. Their model combined acoustoelastic effects and viscoelastic effects and described flexural-wave propagation in pre-stressed materials. However, such models are difficult to invert for mechanical parameter estimation.   

This complexity motivates the use of simpler one-dimensional tension–beam models in elastography, allowing the effects of pre-stretch on transverse-wave propagation to be examined more directly. we use the tensioned Timoshenko model previously studied by \citet{salehabadiQuantifyingUniaxialPrestress2023}.

Building on our previous study~\citep{Zhu2025YoungModulusRod}, the present work extends the investigation to transverse-wave propagation in prestressed soft beams. We employ optical elastography as a high spatio-temporal resolution experimental method to measure the phase-velocity dispersion curves of transverse waves propagating along soft, incompressible, pre-stretched elastic beams. We then assess whether a tensioned beam model can reproduce the observed dependence on both frequency and applied force, and further examine how the individual terms of the model account for the characteristic behavior of the T-wave dispersion curve across different frequency ranges.

\section*{Transverse wave observation}
\begin{figure}[H]
   \centering
   \includegraphics[width=0.8\linewidth]{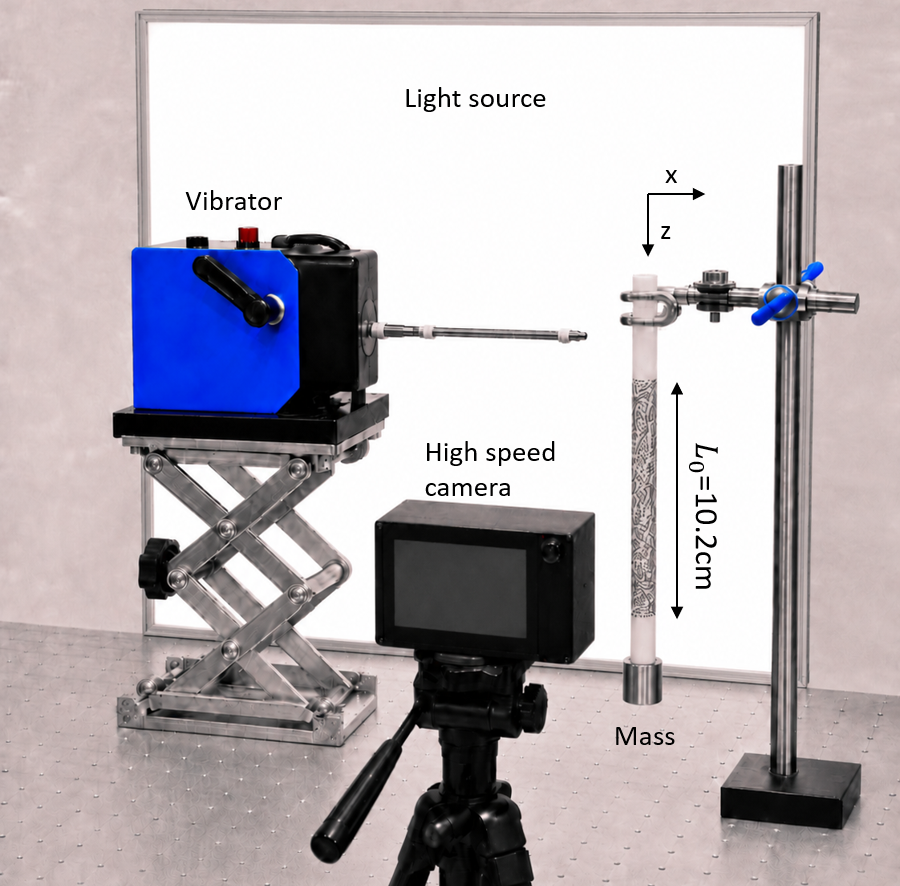}
    \caption{Experimental setup for optical elastography measurements on silicone-based rods. An High-speed camera operating at 3030 Hz was used to image transverse waves propagating along the rod. The waves were generated by a mechanical vibrator, and different masses were applied to control the axial pre-stress in the sample.}
    \label{fig:1}
\end{figure}

The generation and detection of transverse was performed using the  optical elastography setup shown in Figure~\ref{fig:1}. The system consisted of a high-speed camera, a vibrator, and a cylindrical silicone sample. The silicone rod was made of Ecoflex 00-20 Fast (Smooth-On, Macungie, USA), with a density of $\rho = 1070~\mathrm{kg/m^3}$. This material exhibits linear behavior up to 25\% strain. The rod had an initial length of $L = 17.1$ cm and a radius of $r = 3.25$ mm. A central region of length $L_0 = 10.2$ cm was marked with black dots, as shown in Figure~\ref{fig:1}, for motion estimation through Digital Image Correlation. The upper end of the rod was fixed with a clamp. The vibrator (model K2004E01,  The Modal Shop Inc., OH, PA, USA) was connected to a cylinder, and the cylinder was placed in perpendicular contact with the sample to generate transverse waves along the rod. The experiments were performed in air using transmitted-light illumination, with ambient light suppressed.

\begin{figure}[h]
    \centering
    \includegraphics[width=0.9\linewidth]{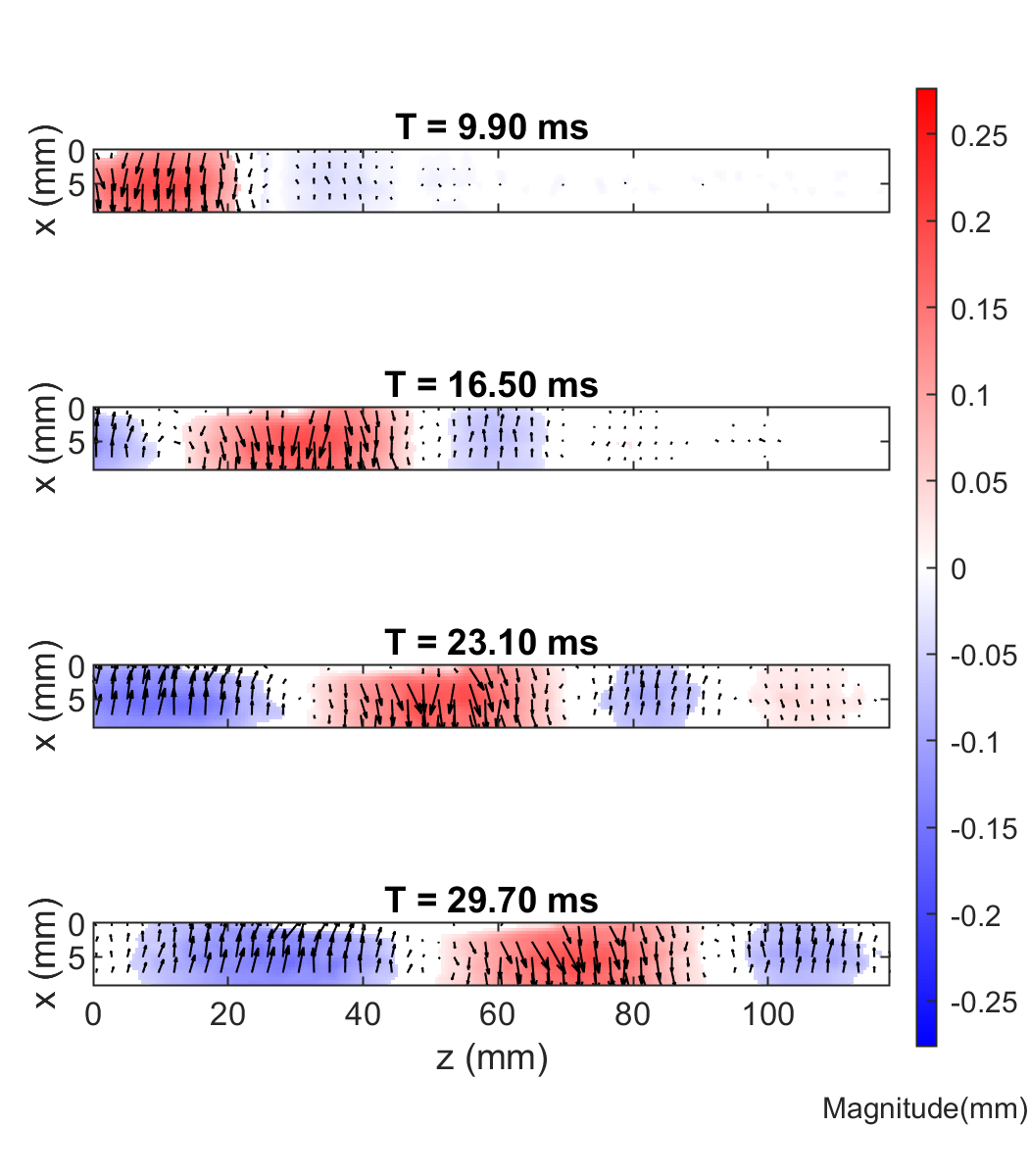}
    \caption{Snapshots of displacements fields taken at different times T for the rod sample subjected to a tensile strain of $\epsilon = 1.9\%$. Arrows describe the direction of the local displacement ; whereas color indicate the amplitude of the vertical displacement.}
    \label{fig:2}
\end{figure}

\begin{figure}[h]
   \centering
   \includegraphics[width=\linewidth]{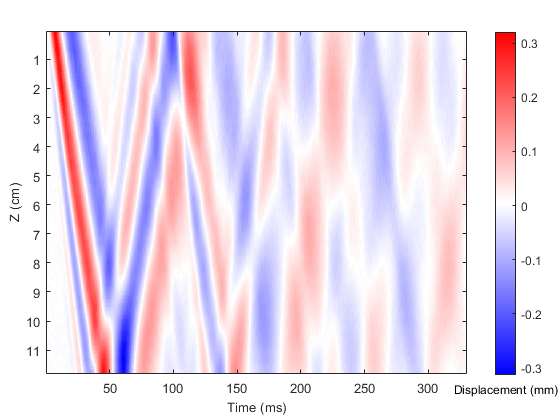}
    \caption{Transverse displacements, averaged along the $x$-axis, detected at each position $z$ over time, for the rod sample subjected to a tensile strain of $\epsilon = 15.6\%$.}
    \label{fig:3}
\end{figure}

Masses ranging from $6$~g to $86.35$~g were attached to the lower end of the rod to apply tension.  The vibrator was then driven using an electrical sinusoidal pulse signal centered at $200$~Hz central frequency. The inertia of the shaker yielded waves at frequencies lower than 200~Hz. After excitation, the wave motion in the marked region was recorded at $3030$~Hz with a resolution of 1280$\times$360 pixels using a high-speed camera system (Chronos 1.4, Kron Technologies, Burnaby, Canada) equipped with a zoom lens (Computar M6Z1212-3S, Computar, Mebane, NC). The recorded video was transferred to a personal computer and processed in MATLAB (MathWorks, Natick, MA). The marked region was selected as the region of interest. Displacements were extracted using a Digital Image Correlation algorithm~\cite{Perrot2018, Garcia2021} with 32$\times$32 pixel correlation windows and 90\% overlap. This produced the displacement field shown in Figure~\ref{fig:2}, where the displacement field is shown with arrows indicating polarization and a color map representing the $x$ component. The wave propagates from the left side to the right side of the sample. 
Since the transverse component is dominant, the displacement field was reduced to a one-dimensional representation by averaging along the $x$-axis, that is, across the diameter. The resulting field emphasizes the displacement component along the $x$ direction associated with transverse-wave propagation. A spatio-temporal representation of this field is shown in Figure~\ref{fig:3}.

\section*{Phase velocity estimation using correlation approaches}
To obtain the phase velocity dispersion curve, we performed a correlation based method that was first used in seismology
\citep{akiQuantitativeSeismology2002} before being adapted for elastography\citep{ganeauCharacterizationViscoelasticProperties2023,Laloy2023}.  The implementation of this approach allows the use of the full recorded wave field displayed in Figure~\ref{fig:3} : the method does not require any directional filtering contrary to time-of-flight methods.  

\begin{figure}[H]
  \centering

    \includegraphics[width=\linewidth]{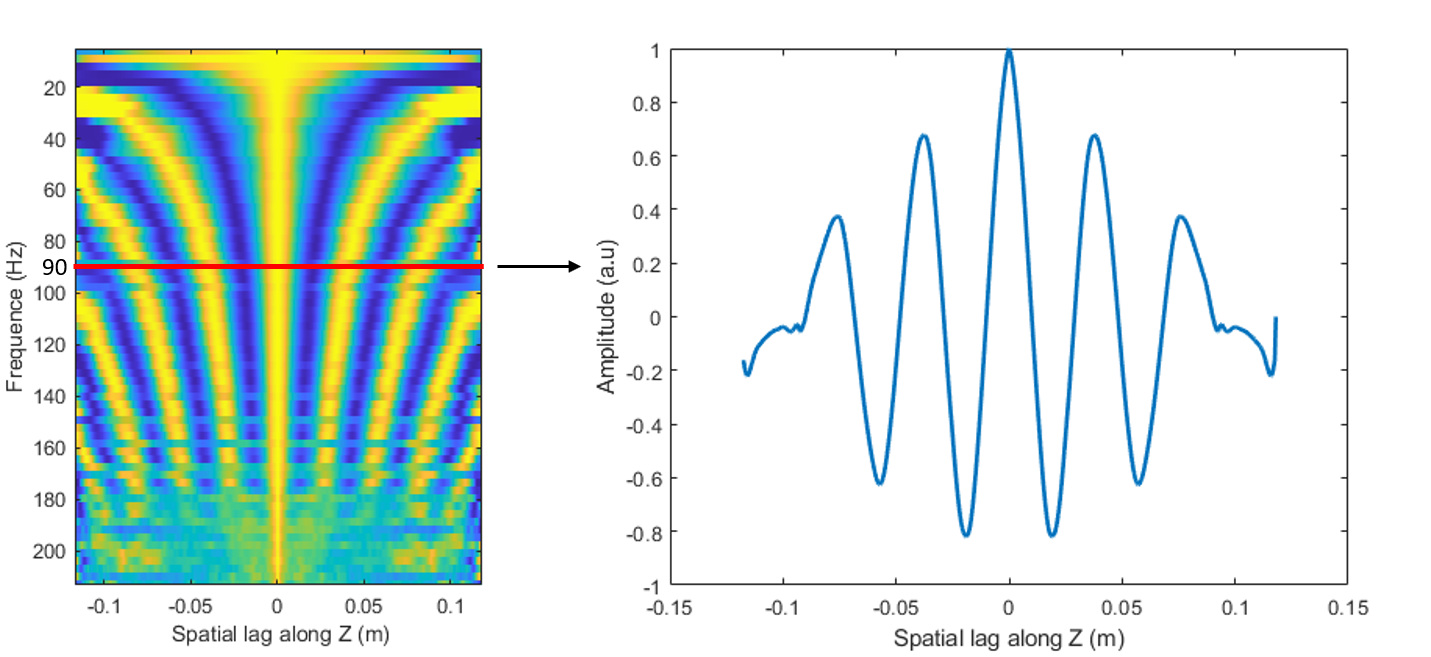} 
  \caption{Spatio-temporal correlation $R(\Delta z ,f)$ for all frequency and at 90 Hz.}
  \label{fig:correlation}
\end{figure}

Given a signal $s(z,t)$  and its temporal Fourier transform $S(z,f)$, for two positions \(z_j\) and \(z_k\), the normalized cross-spectrum is
\begin{equation}
C(z_j,z_k,f)=\frac{S(z_i,f)S^*(z_k,f)}{\mid S(z_i,f)\mid\mid S^*(z_k,f)\mid}.
\end{equation}
The normalization allows only the phase information to be used and is similar to a spectrum whitening process used in seismology for noise correlation techniques \cite{tsarsitalidouLongPeriodRayleigh2024}.
Firstly, considering the difference in position $\Delta z$ such as $z_k=z_i+\Delta z$ ; and  that the medium is homogeneous, we average over the positions $z_i$ yielding:
\begin{equation}
   R(\Delta z,f) =Re\left\langle\frac{S(z_i,f)S^*(z_i+\Delta z,f)}{\mid S(z_i,f)\mid\mid S^*(z_i+\Delta z,f)\mid}\right\rangle_{z_i}.
   \label{eq:C_corr}
\end{equation}
Under the assumption of a random wave field, Equation~\ref{eq:C_corr} is related to the Green's function $G$:
\begin{equation}
   R(\Delta z,f) \propto \operatorname{Im}[G(\Delta z,f)].
   \label{eq:C_Green}
\end{equation}
In our experiment, the transverse wave can be considered a one-dimensional propagating mode. In this case, the spatial dependence of $G$ can be described by a cosine-like oscillation \cite{nakaharaFormulationSpatialAutocorrelation2012}: 
\begin{equation}
   R(\Delta z,f) \propto \cos\!\left[k\Delta z \right] 
   \label{eq:C_1D}
\end{equation}
where $k(f)$ is the wavenumber at the frequency $f$.
Therefore, by computing the cross-correlation for all possible pairs of positions \(z_j\) and \(z_k\) at each frequency, we obtain a frequency-space map, as shown in Figure~\ref{fig:correlation}(a). For each frequency, the spatial variation of the cross-correlation is then fitted with a sinusoidal function to estimate the wavenumber \(k(f)\). The phase velocity is calculated as \(c_\phi(f)=\frac{2\pi f}{k(f)}\). Applying this procedure across the frequency range yields the phase velocity dispersion curve shown in Figure~\ref{fig:6}.

\section*{Elastic parameters estimation using Timoshenko beam theory.}

After plotting the phase velocity dispersion curve under different tension levels, we deployed a Timoshenko-tension coupled model (Eq.~\ref{eq:K}) to set up an inversion method\citep{graff2012wave,salehabadiQuantifyingUniaxialPrestress2023} :
\begin{equation}
    EI\frac{\partial^4 u}{\partial z^4}+T\frac{\partial^2 u}{\partial z^2}+\rho S \frac{\partial^2 u}{\partial t^2}- \rho I \left(1+\frac{E}{\kappa G}\right)\frac{\partial^4 u}{\partial z^2 \partial t^2}+\frac{\rho^2 I}{\kappa G} \frac{\partial^4 u}{\partial t^4}=0.
\label{eq:Timoshenko}
\end{equation}
This model incorporates bending stiffness $EI$, shear deformation $\kappa G$, and axial tension $T$ in a unified framework to capture the competition between bending-dominated and tension-dominated wave behavior. More precisely, $E$ is the Young's modulus, $I$ the second moment of area, $G$ the shear modulus, $\kappa$ the shear coefficient of Timoshenko, $S$ the cross-sectional area, and $\rho$ the density.

In order to obtain the dispersion relation from Equation~\ref{eq:Timoshenko}, we introduce a solution $u(z,t)\propto\exp^{j(kz-\omega t)}$. Equation~\ref{eq:Timoshenko} has four solutions in terms of wave number $k$; and two in terms of $k^2$. We keep the propagative solutions such as $k^2\geq 0$. The other ones correspond to evanescent waves. Therefore, the solutions  are given by:
\begin{equation}
\begin{aligned}
k^2={}&
-\textcolor{TensionOrange}{\frac{T}{2EI}}
+\textcolor{TimoRed}{\frac{\rho\omega^2}{2}
\left(
\frac{1}{E}
+\frac{1}{\kappa G}
\right)}
\\
&+\frac{1}{2}
\sqrt{
\left[
\textcolor{TensionOrange}{\frac{T}{EI}}
-\textcolor{TimoRed}{
\rho\omega^2
\left(
\frac{1}{E}
+\frac{1}{\kappa G}
\right)}
\right]^2
-4\left[
\textcolor{TimoGreen}{
\frac{\rho^2\omega^4}{\kappa GE}
}
-\textcolor{EBblue}{
\frac{\rho S\omega^2}{EI}
}
\right]
}.
\end{aligned}
\label{eq:K}
\end{equation}
The blue term represents the Euler–Bernoulli contribution, while the orange terms account for the applied axial tension. The red terms represent the Timoshenko corrections associated independently with rotary inertia and shear deformation. The green term is a higher-order product-coupling term resulting from the interaction between these two Timoshenko effects.

 For the inversion process, the density $\rho$, cross-sectional area $S$, and second moment of area $I$ were considered known and calculated from each experiment using the incompressibility assumption; thus, a Poisson's ratio of $\nu\approx 0.5$ was used.   Indeed, for a cylindrical incompressible rod, the geometrical parameters vary as :
 \begin{equation}
     \begin{cases}
         r = \frac{r_0}{\sqrt{1+\epsilon}},\\
         I = \frac{1}{4}\pi r^4, \\
         S = \pi r^2.
     \end{cases}
     \label{eq:Geo_param}
 \end{equation}
 
The Young's modulus $E$ and the applied axial tension $T$ were then treated as fitting parameters. Given that the material  was approximately incompressible, the shear modulus $G$ was not fitted independently but was related to the Young's modulus $E$ through $E = 3G$. Only the experimental data below 150~Hz were fitted. Every fitting was performed separately for each tension level. Independent reference values of $E$, and $T$ were obtained from stress-strain experiments \cite{Zhu2025YoungModulusRod} and were used for results comparison and validation. 

\begin{figure}[h]
    \centering
    \includegraphics[width=1\linewidth]{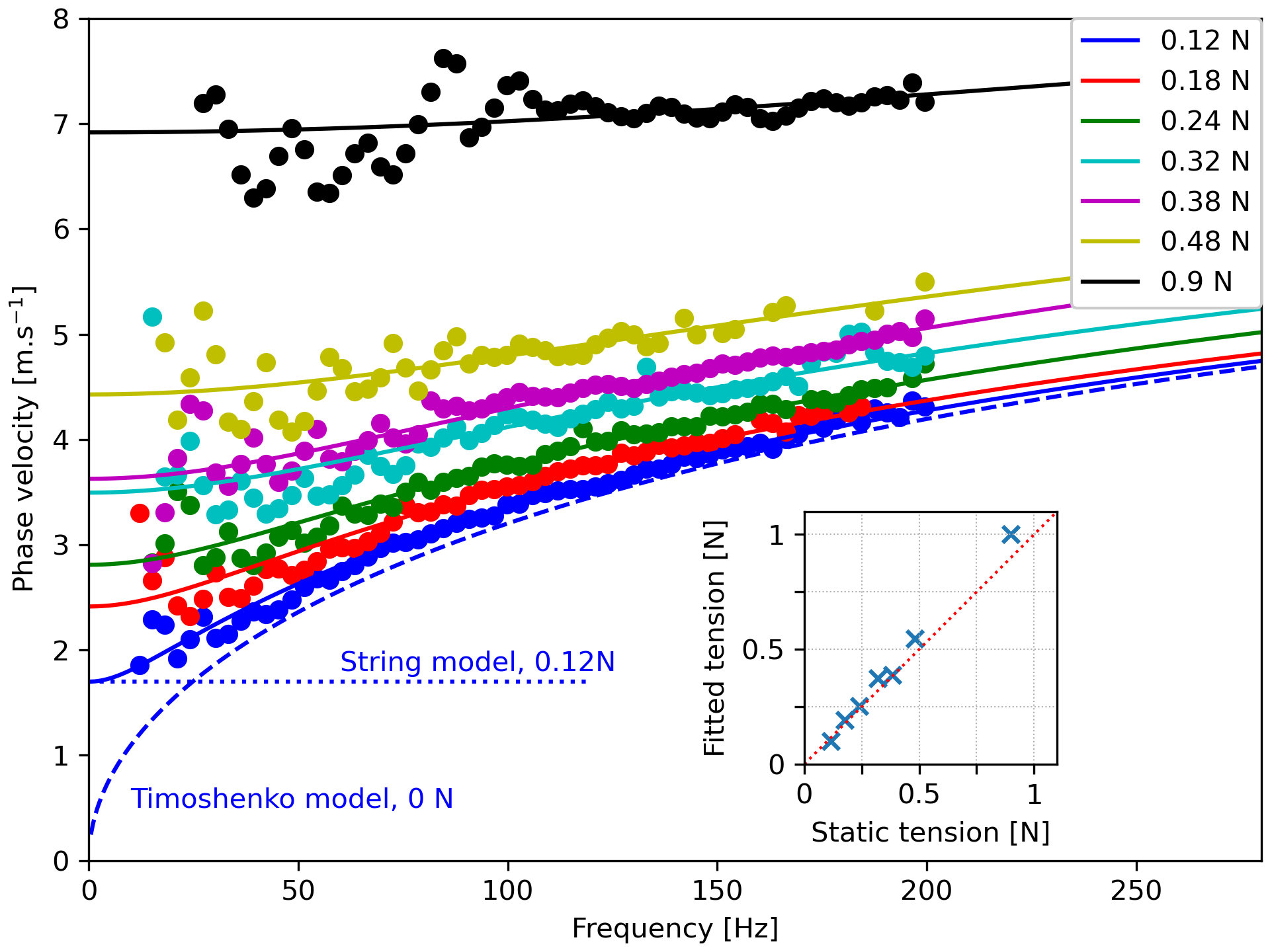}
    \caption{Experimental dispersion curves compared to the fitted model. Different colors are for different tension levels. The scatter points represent the experimental data, the solid lines correspond to the results from the inversion, and the dashed line depicts the pure Timoshenko model velocity. Inset : Tension obtained from the inversion compared to the applied static tension (crosses); the dotted red line corresponds to the equality between both.}
    \label{fig:6}
\end{figure}

The inversion was performed through nonlinear least-squares regressions using the Levenberg-Marquardt algorithm implemented in the \texttt{least\_squares} function from the \texttt{scipy.optimize} Python's module.
The resulting fitted dispersion curves are shown in Fig.~\ref{fig:6} and the estimated  values of the Young's modulus $E$, and the axial tension $T$ are listed in Table~1.

\begin{table}[htbp]
\centering
\caption{Results  of the nonlinear regression for the Young's modulus,  and the tension under different applied tension.  The geometrical parameters are determined by the radius $r$ calculated for each pre-stress through the radius $r=\nicefrac{r_0}{\sqrt{1+\epsilon}}$. The Timoshenko shear coefficient is fixed as $\kappa=9/10$ for cylindrical incompressible materials \cite{cowperShearCoefficientTimoshenkos1966}. As a reminder, the static Young's modulus is estimated as $E_{\mathrm{static}} = 59$ kPa \cite{Zhu2025YoungModulusRod}.}
\label{tab:fitted_results}
\begin{tabular}{ccccc}
\hline
 Applied strain $\epsilon$ (\%)&Applied $T$ (N)&Radius (mm)& Fitted $E$ (kPa)& Fitted  $T$ (N)\\
\hline
 1.9&0.12  &3.22& 155 & 0.10  \\
 6.8&0.18  &3.14& 155 & 0.19  \\
 10.7&0.24  &3.09& 158 & 0.25  \\
 15.6&0.32  &3.02& 176 & 0.38  \\
 20.6&0.38  &2.96& 190 & 0.39  \\
 27.5&0.48  &2.88& 226 & 0.54  \\
 69.6&0.90  &2.5& 231 & 1.00  \\
\hline
\end{tabular}
\end{table}

In the very low-frequency range (0--20~Hz), the tension-wave dominates the wave behavior. In this regime, the model curves remain close to the tension-wave velocity and show, at the highest tensions, a plateau typical of no dispersion. This tension-dominated regime increases with increasing applied tension. As shown in Fig.~\ref{fig:6}, the model captures the phase velocity dispersion curves under different tension levels, and the fitted axial tension values $T$ are consistent with the reference values obtained from the static experiments listed in Table \ref{tab:fitted_results}.

Once the frequency passes a threshold that depends on the tension level, the transverse-wave behavior becomes beam-wave dominated, with clear dispersion appearing due to the bending stiffness, shear deformation, and geometric characteristics of the beam.

However, although the model reproduces the experimental dispersion curves, the Young's modulus $E$ from the fitted values is much larger than the reference value obtained from our stress-strain experiments (Table~\ref{tab:fitted_results}).  Indeed,  considering the conclusion of the previous study~\cite{Zhu2025YoungModulusRod} that the beam material cannot be described without accounting for rheological effects, we should use a complex Young's modulus following a Zener model :  
\begin{equation}
E(\omega) =E_1+\frac{j \omega \eta E_2}{E_2 + j \omega \eta }, 
\label{eq:E}
\end{equation}

Noting that the fitted Young's modulus is of the same order of magnitude as $E_1+E_2$ defined in Eq.~\ref{eq:E}, the fitted value is interpreted here as an effective Young's modulus of the beam. We then reintroduce the Young's modulus defined in Eq.~\ref{eq:E} into the inverse fitting. The results show no essential change is observed in the fitted dispersion curves.

\section*{Discussion}
An ideal string predicts an approximately frequency-independent phase velocity, as illustrated by the dashed points labeled "string model" in Fig .~\ref{fig:6}. In contrast, an untensioned beam exhibits dispersive flexural wave behavior as represented by the dashed line labeled "Timoshenko model" in Fig .~\ref{fig:6}. The simultaneous observation of a tension-dependent velocity and finite-frequency dispersion therefore supports a coupled tension–bending description.

In the present study, we employed a Timoshenko beam model as the final choice for this coupling. Compared with the Euler–Bernoulli model, the Timoshenko model includes not only bending stiffness but also shear deformation and rotary inertia, which are relevant to the propagation of transverse waves and their dispersion characteristics. Since the Timoshenko-based coupled model reproduces the experimental curves within the deformation regime considered here, a more complex model is not required for the purposes of the present study. Furthermore, the fitted tension values agree well with the reference values reported in our previous study \cite{Zhu2025YoungModulusRod}, as shown in the inset in the lower-right corner of Fig.~\ref{fig:6}. This agreement provides additional validation of the coupled model and the fitting procedure.

Revisiting Eq.~\eqref{eq:K} under zero and non-zero axial tension provides a clearer theoretical interpretation of how tension modifies the low-frequency contributions of the Timoshenko model, thereby explaining the experimentally observed behavior.

For T=0, Eq.~\eqref{eq:K} can be rearranged into the normalized form given in Eq.~\eqref{eq:K_T0} : 
\begin{equation}
\frac{k^2}{\omega^2}
=
\textcolor{TimoRed}{
\frac{\rho}{2}
\left(
\frac{1}{E}
+\frac{1}{\kappa G}
\right)}
+
\sqrt{
\textcolor{EBblue}{
\frac{\rho S}{EI\omega^2}
}
+
\textcolor{TimoRed}{
\frac{\rho^2}{4}
\left(
\frac{1}{E}
-\frac{1}{\kappa G}
\right)^2
}
}.
\label{eq:K_T0}
\end{equation}

Expanding Eq.~\eqref{eq:K_T0} in the low-frequency limit $\omega\rightarrow 0$ up to and including terms of order $\omega^3$ gives :
\begin{equation}
    k^2=\sqrt{\frac{\rho S}{E I}}\omega+\frac{1}{2}\rho \omega^2 \left(\frac{1}{E}+\frac{1}{\kappa G}\right) + \frac{1}{8}\rho ^2\omega^3\sqrt{\frac{ E I}{\rho S}}\left(\frac{1}{E}-\frac{1}{\kappa G}\right)^2 + \mathcal{O}(\omega^3)
    \label{eq:K_T0_dl}
\end{equation}
This expansion makes the hierarchy of the different physical contributions more transparent. In the absence of axial tension, the leading term scales as $\mathcal{O}(\omega)$ and corresponds to the Euler--Bernoulli flexural-wave behavior. The contribution associated with shear deformation and rotary inertia first enters at $\mathcal{O}(\omega^2)$, with additional corrections appearing at higher orders. These effects therefore modify the finite-frequency dispersion but do not determine the leading low-frequency behavior when $T=0$.

For $T\neq0$, Eq.~\eqref{eq:K} can be rearranged to make the frequency-dependent terms explicit:

\begin{equation}
\begin{aligned}
k^2={}&
\textcolor{TensionOrange}{
-\frac{T}{2EI}
}
+
\textcolor{TimoRed}{
\frac{\rho\omega^2}{2}
\left(
\frac{1}{E}
+\frac{1}{\kappa G}
\right)}
\\
&+
\frac{1}{2}
\sqrt{
\textcolor{TensionOrange}{
\left(
\frac{T}{EI}
\right)^2
}
+
\textcolor{EBblue}{
\omega^2
\left[
\frac{4\rho S}{EI}
-
\frac{2\rho\,\textcolor{TensionGreen}{T}}{EI}
\left(
\frac{1}{E}
+\frac{1}{\kappa G}
\right)
\right]}
+
\textcolor{TimoRed}{
\rho^2\omega^4
\left(
\frac{1}{E}
-\frac{1}{\kappa G}
\right)^2
}
}.
\end{aligned}
\label{eq:K_T_nonzero}
\end{equation}

Performing a series expansion  of Equation~\ref{eq:K_T_nonzero} for $\omega\rightarrow0$ up to order 4 yields 
\begin{equation}
    k^2=\textcolor{TensionGreen}{\frac{\rho S}{T}}\omega^2+\rho^2\omega^4 \left[-\frac{I}{T\kappa G}+\frac{ I S}{T^2}\left(1+\frac{E}{\kappa G}\right)-\frac{EIS^2}{T^3}\right]+\mathcal{O}(\omega^4).
    \label{eq:K_T_nonzero_DL_4}
\end{equation}
The first term of Equation~\ref{eq:K_T_nonzero_DL_4}, displayed in green explicitly depends on the applied tension. The corresponding  phase velocity  is 
\begin{equation}
    c_\phi=\omega/k\approx\sqrt{\frac{T}{\rho S}},  
    \label{eq:tension_speed}
\end{equation}
confirming that the model correctly predicts the behavior of string waves at low frequencies. 
The next term in Equation~\ref{eq:K_T_nonzero_DL_4} depends on $\omega^4$. This development demonstrates that the low-frequency term of Euler-Bernoulli, proportional to $\omega$, as in Equation~\ref{eq:K_T0}, has disappeared. The  tension becomes the principal force and  impacts the description of all the dispersion curves.  

In the high-frequency limit, $\omega \rightarrow \infty$, both Eq.~\eqref{eq:K_T_nonzero} for $T\neq 0$ and Eq.~\eqref{eq:K_T0} for $T=0$ give the same leading-order expression,
\begin{equation}
    k^2 \sim \frac{\rho \omega^2}{\kappa G}.
    \label{eq:K_omegainfini}
\end{equation}
Therefore,
\begin{equation}
    c_\phi=\frac{\omega}{k}
    \longrightarrow
    \sqrt{\frac{\kappa G}{\rho}}.
    \label{eq:flexion_speed}
\end{equation}
This result shows that the effect of axial tension becomes negligible in the high-frequency limit. The high-frequency behavior is instead governed by shear deformation, and the phase velocity of the flexural wave approaches the effective shear-wave velocity $\sqrt{\kappa G/\rho}$ predicted by the Timoshenko beam model.

These asymptotic results support a broader physical interpretation within the class of tensioned beam-like waveguides considered here. Once axial tension contributes appreciably to transverse-wave propagation, the measured behavior can no longer be interpreted as arising from flexural rigidity alone. Instead, it reflects the coupled influence of axial tension and beam-related deformation mechanisms, with their relative importance depending on frequency, loading, geometry, and material properties.

A particularly relevant practical implication concerns transverse-wave elastography of tendons and other slender, fiber-reinforced soft tissues subjected to axial loading. Although tendon architecture and constitutive behavior are considerably more complex than those of the homogeneous cylindrical specimens investigated here, tendons commonly support substantial tensile loads. Consequently, elastic moduli inferred from transverse-wave measurements without explicitly accounting for pre-stress may contain tension-induced contributions. Such parameters should therefore be interpreted as effective wave-based quantities rather than being identified directly with the intrinsic or quasi-static material modulus. The magnitude of this effect, and whether prestress and intrinsic elasticity can be independently recovered in biological tissues, were not quantitatively evaluated in the present study.

Similar ideas have been proposed in previous studies. In 2021, Zhao and Chang~\cite{ZHAO2021111208} showed theoretically and numerically that for transverse waves in finitely pre-stretched soft fibers under relatively small pre-strain, a Rayleigh beam model incorporating axial tension was required to describe the increase in wave velocity with increasing applied tension. Their study established the theoretical relevance of bending–tension coupling, whereas direct validation against experimentally measured phase-velocity dispersion curves was outside its scope. The present experiments complement this earlier work by testing the coupled description directly over the measured frequency range.

More recently, Delory et al.~\cite{D3SM01485A} applied a similar experimental methodology to investigate guided waves in a highly stretched strip. Their results likewise showed that the transverse-wave velocity increased with the applied tension, providing further support for the tension-dependent behavior observed in the present study. They interpreted this coupled wave behavior using a more comprehensive framework incorporating acoustoelasticity, nonlinear hyperelasticity, and viscoelasticity. This framework successfully reproduced their experimental dispersion curves and provided a detailed description of the material response. However, in their model, the individual physical contributions are not clearly separated, making their respective roles in the measured dispersion less direct and intuitive. By contrast, the reduced-order model used in the present study introduces axial tension and beam rigidity explicitly as distinct mechanical terms. Their different dependencies on wavenumber and frequency make it possible to identify directly how tension governs the low-frequency string-like response, whereas bending stiffness, shear deformation, and rotary inertia control the finite-frequency dispersion. The present model, therefore, does not merely provide an alternative fit to the experimental curves; it exposes the separate physical mechanisms underlying the observed bending–tension coupling.

Regarding viscoelasticity, our previous longitudinal-wave study on the same sample under comparable tension conditions showed that the material rheology could be described using a Zener model~\cite{Zhu2025YoungModulusRod}. Introducing the same frequency-dependent complex modulus into the tensioned-Timoshenko model produced only a limited change in the transverse-wave dispersion, with differences below approximately 5 $\%$ at low frequencies and becoming nearly negligible at higher frequencies. This suggests that the observed high-frequency dispersion is mainly governed by beam-waveguide effects rather than viscosity. However, the Young's modulus obtained from the purely elastic fitting was close to the rheological-model values at lower tension levels, whereas larger deviations appeared as the applied tension increased, with fitted values reaching approximately 231 kPa. Therefore, although viscosity has little influence on the dispersion curve itself, the fitted Young's modulus should still be interpreted as an effective dynamic parameter rather than directly as the static elastic modulus.

\section*{Conclusion}

In summary, the coupled tension-Timoshenko beam model reproduced the transverse-wave dispersion curves measured in the soft cylindrical beam under different axial tension levels. The model provides a clear physical interpretation of the measured dispersion. At low frequencies, the wave behavior is mainly controlled by axial tension. At higher frequencies, bending, shear deformation, and rotary inertia become more important. These different regimes predicted by the model can be identified directly in the experimental dispersion curves over the investigated frequency and loading ranges.

The model provides a physically interpretable framework for studying transverse wave propagation in beam-like structures under axial tension. It may also be useful for characterizing similar pre-stretched structures in practical applications. Our next step will be to apply this model to tendons and examine whether their effective elastic properties can be estimated from transverse wave dispersion measurements while accounting for the tension effect.

\section*{CONFLICT OF INTEREST STATEMENT}
The authors declare that they have no financial interests, commercial affiliations, or other potential conflicts of interest that could have influenced the objectivity of this research or the writing of this paper.

\section*{CODE, DATA, AND MATERIALS AVAILABILITY}
Data and codes are available upon reasonable request to the corresponding author.

\section*{ACKNOWLEDGMENTS}
 With financial support from ITMO Cancer of Aviesan within the framework of the 2021-2030 Cancer Control Strategy, on funds administered by Inserm.
\bibliographystyle{achemso}
\bibliography{biblio}

\providecommand{\latin}[1]{#1}
\makeatletter
\providecommand{\doi}
  {\begingroup\let\do\@makeother\dospecials
  \catcode`\{=1 \catcode`\}=2 \doi@aux}
\providecommand{\doi@aux}[1]{\endgroup\texttt{#1}}
\makeatother
\providecommand*\mcitethebibliography{\thebibliography}
\csname @ifundefined\endcsname{endmcitethebibliography}  {\let\endmcitethebibliography\endthebibliography}{}
\begin{mcitethebibliography}{19}
\providecommand*\natexlab[1]{#1}
\providecommand*\mciteSetBstSublistMode[1]{}
\providecommand*\mciteSetBstMaxWidthForm[2]{}
\providecommand*\mciteBstWouldAddEndPuncttrue
  {\def\EndOfBibitem{\unskip.}}
\providecommand*\mciteBstWouldAddEndPunctfalse
  {\let\EndOfBibitem\relax}
\providecommand*\mciteSetBstMidEndSepPunct[3]{}
\providecommand*\mciteSetBstSublistLabelBeginEnd[3]{}
\providecommand*\EndOfBibitem{}
\mciteSetBstSublistMode{f}
\mciteSetBstMaxWidthForm{subitem}{(\alph{mcitesubitemcount})}
\mciteSetBstSublistLabelBeginEnd
  {\mcitemaxwidthsubitemform\space}
  {\relax}
  {\relax}

\bibitem[Martin \latin{et~al.}(2018)Martin, Brandon, Keuler, Hermus, Ehlers, Segalman, Allen, and Thelen]{Martin2018GaugingForceTendons}
Martin,~J.~A.; Brandon,~S. C.~E.; Keuler,~E.~M.; Hermus,~J.~R.; Ehlers,~A.~C.; Segalman,~D.~J.; Allen,~M.~S.; Thelen,~D.~G. Gauging force by tapping tendons. \emph{Nat. Commun.} \textbf{2018}, \emph{9}, 1592\relax
\mciteBstWouldAddEndPuncttrue
\mciteSetBstMidEndSepPunct{\mcitedefaultmidpunct}
{\mcitedefaultendpunct}{\mcitedefaultseppunct}\relax
\EndOfBibitem
\bibitem[Keuler \latin{et~al.}(2019)Keuler, Loegering, Martin, Roth, and Thelen]{Keuler2019AchillesTendonLoading}
Keuler,~E.~M.; Loegering,~I.~F.; Martin,~J.~A.; Roth,~J.~D.; Thelen,~D.~G. Shear Wave Predictions of Achilles Tendon Loading during Human Walking. \emph{Scientific Reports} \textbf{2019}, \emph{9}, 13419\relax
\mciteBstWouldAddEndPuncttrue
\mciteSetBstMidEndSepPunct{\mcitedefaultmidpunct}
{\mcitedefaultendpunct}{\mcitedefaultseppunct}\relax
\EndOfBibitem
\bibitem[Rougereau \latin{et~al.}(2022)Rougereau, Marty-Diloy, Vigan, Donadieu, Hardy, Vialle, and Langlais]{Rougereau2022DeltoidLigamentSWE}
Rougereau,~G.; Marty-Diloy,~T.; Vigan,~M.; Donadieu,~K.; Hardy,~A.; Vialle,~R.; Langlais,~T. A Preliminary Study to Assess the Relevance of Shear-Wave Elastography in Characterizing Biomechanical Changes in the Deltoid Ligament Complex in Relation to Ankle Position. \emph{Foot \& Ankle International} \textbf{2022}, \emph{43}, 840--849\relax
\mciteBstWouldAddEndPuncttrue
\mciteSetBstMidEndSepPunct{\mcitedefaultmidpunct}
{\mcitedefaultendpunct}{\mcitedefaultseppunct}\relax
\EndOfBibitem
\bibitem[Yancey \latin{et~al.}(2023)Yancey, Rbil, Chatterjee, Lin, Wyles, Ko, Nwawka, and Khormaee]{Yancey2023SpinalLigamentSWE}
Yancey,~M.; Rbil,~N.; Chatterjee,~A.; Lin,~H.; Wyles,~H.~L.; Ko,~L.~M.; Nwawka,~O.~K.; Khormaee,~S. Ultrasound Shear Wave Elastography Quantitatively Assesses Tension Changes of Supraspinous/Interspinous Ligament Complex Under Varied Loads. \emph{International Journal of Spine Surgery} \textbf{2023}, \emph{17}, 502--510\relax
\mciteBstWouldAddEndPuncttrue
\mciteSetBstMidEndSepPunct{\mcitedefaultmidpunct}
{\mcitedefaultendpunct}{\mcitedefaultseppunct}\relax
\EndOfBibitem
\bibitem[Zhao and Chang(2021)Zhao, and Chang]{ZHAO2021111208}
Zhao,~S.; Chang,~Z. Elastic wave velocities in finitely pre-stretched soft fibers. \emph{Int. J. Solids Struct.} \textbf{2021}, \emph{233}, 111208\relax
\mciteBstWouldAddEndPuncttrue
\mciteSetBstMidEndSepPunct{\mcitedefaultmidpunct}
{\mcitedefaultendpunct}{\mcitedefaultseppunct}\relax
\EndOfBibitem
\bibitem[Graff(2012)]{graff2012wave}
Graff,~K.~F. \emph{Wave motion in elastic solids}; Courier Corporation, 2012\relax
\mciteBstWouldAddEndPuncttrue
\mciteSetBstMidEndSepPunct{\mcitedefaultmidpunct}
{\mcitedefaultendpunct}{\mcitedefaultseppunct}\relax
\EndOfBibitem
\bibitem[Salehabadi \latin{et~al.}(2023)Salehabadi, Nammari, Luna, Crutison, Klatt, and Royston]{salehabadiQuantifyingUniaxialPrestress2023}
Salehabadi,~M.; Nammari,~L.; Luna,~A.; Crutison,~J.; Klatt,~D.; Royston,~T.~J. Quantifying Uniaxial Prestress and Waveguide Effects on Dynamic Elastography Estimates for a Cylindrical Rod. \emph{J. Acoust. Soc. Am.} \textbf{2023}, \emph{154}, 3580--3594\relax
\mciteBstWouldAddEndPuncttrue
\mciteSetBstMidEndSepPunct{\mcitedefaultmidpunct}
{\mcitedefaultendpunct}{\mcitedefaultseppunct}\relax
\EndOfBibitem
\bibitem[Ogden(1997)]{ogdenNonlinearElasticDeformations1997}
Ogden,~R.~W. \emph{Non-Linear Elastic Deformations}, 1st ed.; Dover Books on Physics; Dover Publ: Mineola, NY, 1997\relax
\mciteBstWouldAddEndPuncttrue
\mciteSetBstMidEndSepPunct{\mcitedefaultmidpunct}
{\mcitedefaultendpunct}{\mcitedefaultseppunct}\relax
\EndOfBibitem
\bibitem[Delory \latin{et~al.}(2023)Delory, Lemoult, Eddi, and Prada]{Delory2023HighlyStretchedPlate}
Delory,~A.; Lemoult,~F.; Eddi,~A.; Prada,~C. Guided elastic waves in a highly-stretched soft plate. \emph{Extreme Mech. Lett.} \textbf{2023}, \emph{61}, 102018\relax
\mciteBstWouldAddEndPuncttrue
\mciteSetBstMidEndSepPunct{\mcitedefaultmidpunct}
{\mcitedefaultendpunct}{\mcitedefaultseppunct}\relax
\EndOfBibitem
\bibitem[Zhu \latin{et~al.}(2025)Zhu, Legrand, Gr{\'e}goire, Torres, Catheline, and Giammarinaro]{Zhu2025YoungModulusRod}
Zhu,~J.; Legrand,~F.; Gr{\'e}goire,~S.; Torres,~J.; Catheline,~S.; Giammarinaro,~B. Young's modulus estimation of a soft viscoelastic rod using optical elastography. \emph{J. Acoust. Soc. Am.} \textbf{2025}, \emph{158}, 3203--3209\relax
\mciteBstWouldAddEndPuncttrue
\mciteSetBstMidEndSepPunct{\mcitedefaultmidpunct}
{\mcitedefaultendpunct}{\mcitedefaultseppunct}\relax
\EndOfBibitem
\bibitem[Perrot and Garcia(2018)Perrot, and Garcia]{Perrot2018}
Perrot,~V.; Garcia,~D. Back to Basics in Ultrasound Velocimetry: Tracking Speckles by Using a Standard Piv Algorithm. 2018 IEEE International Ultrasonics Symposium (IUS). 2018; pp 206--212\relax
\mciteBstWouldAddEndPuncttrue
\mciteSetBstMidEndSepPunct{\mcitedefaultmidpunct}
{\mcitedefaultendpunct}{\mcitedefaultseppunct}\relax
\EndOfBibitem
\bibitem[Garcia(2021)]{Garcia2021}
Garcia,~D. Make the most of MUST, an open-source Matlab UltraSound Toolbox. 2021 IEEE International Ultrasonics Symposium (IUS). 2021; pp 1--4\relax
\mciteBstWouldAddEndPuncttrue
\mciteSetBstMidEndSepPunct{\mcitedefaultmidpunct}
{\mcitedefaultendpunct}{\mcitedefaultseppunct}\relax
\EndOfBibitem
\bibitem[Aki and Richards(2002)Aki, and Richards]{akiQuantitativeSeismology2002}
Aki,~K.; Richards,~P. \emph{Quantitative {{Seismology}}}; CA: University Science Books: San Francisco, 2002\relax
\mciteBstWouldAddEndPuncttrue
\mciteSetBstMidEndSepPunct{\mcitedefaultmidpunct}
{\mcitedefaultendpunct}{\mcitedefaultseppunct}\relax
\EndOfBibitem
\bibitem[Ganeau \latin{et~al.}(2023)Ganeau, Lafond, Legrand, {Laloy-Borgna}, Ben~Moussa, Poinard, Mascarelli, Thuret, Gain, Lafon, and Catheline]{ganeauCharacterizationViscoelasticProperties2023}
Ganeau,~A.; Lafond,~M.; Legrand,~F.; {Laloy-Borgna},~G.; Ben~Moussa,~O.; Poinard,~S.; Mascarelli,~F.; Thuret,~G.; Gain,~P.; Lafon,~C.; Catheline,~S. Characterization of the Viscoelastic Properties of {\emph{in Vitro}} Crystalline Lens Samples Using Ultrasound Elastography. \emph{Appl. Phys. Lett.} \textbf{2023}, \emph{123}, 103701\relax
\mciteBstWouldAddEndPuncttrue
\mciteSetBstMidEndSepPunct{\mcitedefaultmidpunct}
{\mcitedefaultendpunct}{\mcitedefaultseppunct}\relax
\EndOfBibitem
\bibitem[Laloy-Borgna \latin{et~al.}(2023)Laloy-Borgna, Puyo, Nishino, Atlan, and Catheline]{Laloy2023}
Laloy-Borgna,~G.; Puyo,~L.; Nishino,~H.; Atlan,~M.; Catheline,~S. Observation of natural flexural pulse waves in retinal and carotid arteries for wall elasticity estimation. \emph{Science Advances} \textbf{2023}, \emph{9}, eadf1783\relax
\mciteBstWouldAddEndPuncttrue
\mciteSetBstMidEndSepPunct{\mcitedefaultmidpunct}
{\mcitedefaultendpunct}{\mcitedefaultseppunct}\relax
\EndOfBibitem
\bibitem[Tsarsitalidou \latin{et~al.}(2024)Tsarsitalidou, Hillers, Giammarinaro, Bou{\'e}, Stehly, and Campillo]{tsarsitalidouLongPeriodRayleigh2024}
Tsarsitalidou,~C.; Hillers,~G.; Giammarinaro,~B.; Bou{\'e},~P.; Stehly,~L.; Campillo,~M. Long {{Period Rayleigh Wave Focal Spot Imaging Applied}} to {{USArray Data}}. \emph{Journal of Geophysical Research: Solid Earth} \textbf{2024}, \emph{129}, e2023JB027417\relax
\mciteBstWouldAddEndPuncttrue
\mciteSetBstMidEndSepPunct{\mcitedefaultmidpunct}
{\mcitedefaultendpunct}{\mcitedefaultseppunct}\relax
\EndOfBibitem
\bibitem[Nakahara(2012)]{nakaharaFormulationSpatialAutocorrelation2012}
Nakahara,~H. Formulation of the Spatial Autocorrelation ({{SPAC}}) Method in Dissipative Media. \emph{Geophys. J. Int.} \textbf{2012}, \emph{190}, 1777--1783\relax
\mciteBstWouldAddEndPuncttrue
\mciteSetBstMidEndSepPunct{\mcitedefaultmidpunct}
{\mcitedefaultendpunct}{\mcitedefaultseppunct}\relax
\EndOfBibitem
\bibitem[Cowper(1966)]{cowperShearCoefficientTimoshenkos1966}
Cowper,~G.~R. The {{Shear Coefficient}} in {{Timoshenko}}'s {{Beam Theory}}. \emph{Journal of Applied Mechanics} \textbf{1966}, \emph{33}, 335--340\relax
\mciteBstWouldAddEndPuncttrue
\mciteSetBstMidEndSepPunct{\mcitedefaultmidpunct}
{\mcitedefaultendpunct}{\mcitedefaultseppunct}\relax
\EndOfBibitem
\bibitem[Delory \latin{et~al.}(2024)Delory, Kiefer, Lanoy, Eddi, Prada, and Lemoult]{D3SM01485A}
Delory,~A.; Kiefer,~D.~A.; Lanoy,~M.; Eddi,~A.; Prada,~C.; Lemoult,~F. Viscoelastic dynamics of a soft strip subject to a large deformation. \emph{Soft Matter} \textbf{2024}, \emph{20}, 1983--1995\relax
\mciteBstWouldAddEndPuncttrue
\mciteSetBstMidEndSepPunct{\mcitedefaultmidpunct}
{\mcitedefaultendpunct}{\mcitedefaultseppunct}\relax
\EndOfBibitem
\end{mcitethebibliography}

\end{document}